\documentclass{espcrc2}
\newcommand{\Dstar}{D^{\ast}}
\newcommand{\Fcharm}{F_2^{c\overline{c}}}
\title{Charm production at HERA}
\author{G.P.\ Heath\address{H.H.\ Wills Physics Laboratory, University
of Bristol, Bristol BS8 1TL, UK}\thanks{on behalf of the ZEUS and H1
collaborations.}}
\begin{document}
\begin{abstract}
\end{abstract}
\maketitle

\section{INTRODUCTION}
The HERA collider circulates beams of electrons (positrons) and protons
at energies of 27.5 and $820\,\mbox{GeV}$ respectively. The $e^{\pm}p$
centre of mass energy for collisions between the beams is
$300\,\mbox{GeV}$. The largest cross-section processes observed in these
collisions are photoproduction reactions, where  the $e^{\pm}$ beam acts
as a source of almost real ($Q^2 \ll 1 \,\mbox{GeV}^2$) photons.
Processes where a virtual photon of sizeable $Q^2$ is exchanged are
known as Deep Inelastic Scattering, or DIS reactions. In this paper I
review measurements of open charm production at HERA made in both
photoproduction and DIS by the H1 and ZEUS collaborations.

\begin{figure}[htb]\centering
\epsfig{file=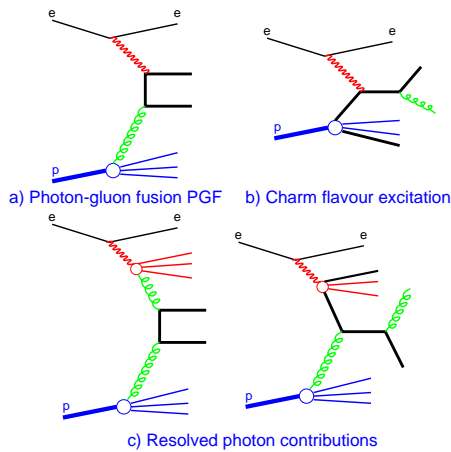}
\caption{\label{fig-Feyn}
Charm production processes at HERA.}
\end{figure}
Charm quarks are produced in real or virtual photon--proton collisions
predominantly via the photon--gluon fusion process illustrated in
Figure~\ref{fig-Feyn}a. Charm production is therefore sensitive to the
gluon momentum distribution in the proton. Other contributing processes
are also shown in Figure~\ref{fig-Feyn}. The diagram of
Figure~\ref{fig-Feyn}b, where the charm quark is an active constituent
of the proton, is expected to be important in DIS at large $Q^2$. In
photoproduction, additional contributions may arise from ``resolved
photon'' processes as shown in Figure~\ref{fig-Feyn}c, which reveal the
partonic structure of the photon. Sources of charm production other than
via the hard processes shown, such as fragmentation or the decay of $b$
quarks, are negligible by comparison.

Cross-section calculations for these processes in the framework of
perturbative QCD should be reliable, since the relatively large charm
quark mass $m_c > \Lambda_{\mbox\tiny{QCD}}$ provides a hard scale in
all cases. In addition, in many kinematic regions, further hard scales
are provided by a large photon virtuality $Q^2$ or high transverse
momentum of the produced charm quark pair. A number of authors have
produced calculations of charm production to next--to--leading order
(NLO) in QCD.

The analyses reviewed here are based on data taken in 1994, when each
experiment collected around $3\,pb^{-1}$ of positron--proton collisions.
Both H1 and ZEUS reconstruct charm signals using the invariant masses of
tracks found in the tracking detectors. No particle identification is
used. I cover only the production of open charm; $c\overline{c}$
meson signals are discussed in the contribution to this conference on
vector mesons at HERA\cite{bib-Meyer}.

Throughout this paper I use standard symbols for event--related
kinematic quantities at HERA. $Q^2$ is the invariant squared 4--momentum
transfer from the positron to the hadronic system, written as a positive
quantity; $Q^2=-q^2=-(k-k')^2$, where $k$, $k'$ are the initial and
final state positron 4--momenta. The Bjorken scaling variable $x$ is
defined by $x = \frac{Q^2}{2p \cdot q}$, where $p$ is the initial proton
beam 4--momentum. $x$ is the fractional momentum carried by the struck
parton, in the parton model of DIS processes. $W$ is the invariant mass
of the final state hadron system, or the $\gamma^{\ast}p$ invariant mass
given by $W^2 = (p+q)^2$. These three quantities are related by $Q^2 =
W^2\left(\frac{1-x}{x}\right)$. Finally, the variable $y=\frac{p \cdot
q}{p \cdot k}=\frac{Q^2}{xs}$ where $\sqrt{s}$ is the total $ep$
centre-of-mass energy.

This review is divided into two main sections, and a short summary. In
section~\ref{sect-gprod} I discuss charm photoproduction measurements
and QCD calculations. Charm production in DIS, and measurements of the
charm contribution to the proton structure function, are presented in
section~\ref{sect-DIS}.

\section{OPEN CHARM IN PHOTOPRODUCTION}
\label{sect-gprod}

The techniques used by the two
collaborations\cite{bib-H1-gprod,bib-ZEUS-gprod} to reconstruct charm
candidates in photoproduction events are similar. The aim is to identify
the decay chain
\begin{equation} \Dstar(2010)^+ \longrightarrow D^0 \pi_s^+
\longrightarrow K^- \left( n\pi \right)^+ \pi_s^+
\label{eq-Dstdecay}\end{equation}
and its charge conjugate, where $\pi_s^+$ is a ``slow'' pion carrying
only $40\,\mbox{MeV}$ of momentum in the ${\Dstar}^+$ rest frame. The
analysis uses pairs of oppositely charged tracks to search for $D^0$
candidates, and ZEUS also has a signal from 4-track combinations. Each
track in turn is assigned the charged kaon mass, with the remainder
assumed to be pions. 
\begin{figure}[htb]\centering
\epsfig{file=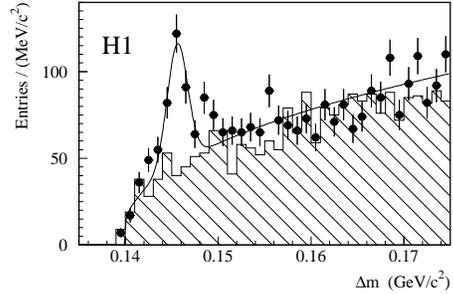,scale=0.38}
\caption{\label{fig-DeltaMdist}
Distribution of $\Delta m$ for $\Dstar$ candidates in photoproduction,
as measured by H1.}
\end{figure}
Combinations whose invariant mass $m_{K\pi}$ falls within a window
around the $D^0$ mass are then further combined with a track of opposite
charge to the kaon candidate, assumed to be the slow pion. The number of
${\Dstar}^{\pm}$ decays in the sample is estimated from the distribution
of $\Delta m = m_{K\pi\pi_s}-m_{K\pi}$. The mass difference distribution
from H1 is shown in Figure~\ref{fig-DeltaMdist}, and a clear signal can
be seen with a relatively small background.

Kinematic cuts are applied to the $\Dstar$ candidates, to ensure clean
reconstruction and reduce background. Both experiments require their
$\Dstar$s to be in the central region of pseudorapidity, $-1.5 < \eta <
1.0$ ($40^{\circ} < \theta < 155^{\circ}$). H1 apply a $p_t$ cut of
$2.5\,\mbox{GeV}$, while ZEUS apply cuts of $3(4)\,\mbox{GeV}$ for
$K\pi\pi_s$ ($K\pi\pi\pi\pi_s$) signals.

Photoproduction event samples with $Q^2<4\,\mbox{GeV}^2$ and $W$ in the
range 100 to $200\,\mbox{GeV}$ are identified by activity in the main
detector, with no scattered positron detected. The average value of
$Q^2$ is around $0.2\,\mbox{GeV}^2$, and the median value $5 \cdot
10^{-4}\,\mbox{GeV}^2$. H1 additionally use a tagged sample where the
positron is detected in the small angle luminosity detector, giving
an average $Q^2$ of $10^{-3}\,\mbox{GeV}^2$.

\begin{figure}[htb]\centering
\epsfig{file=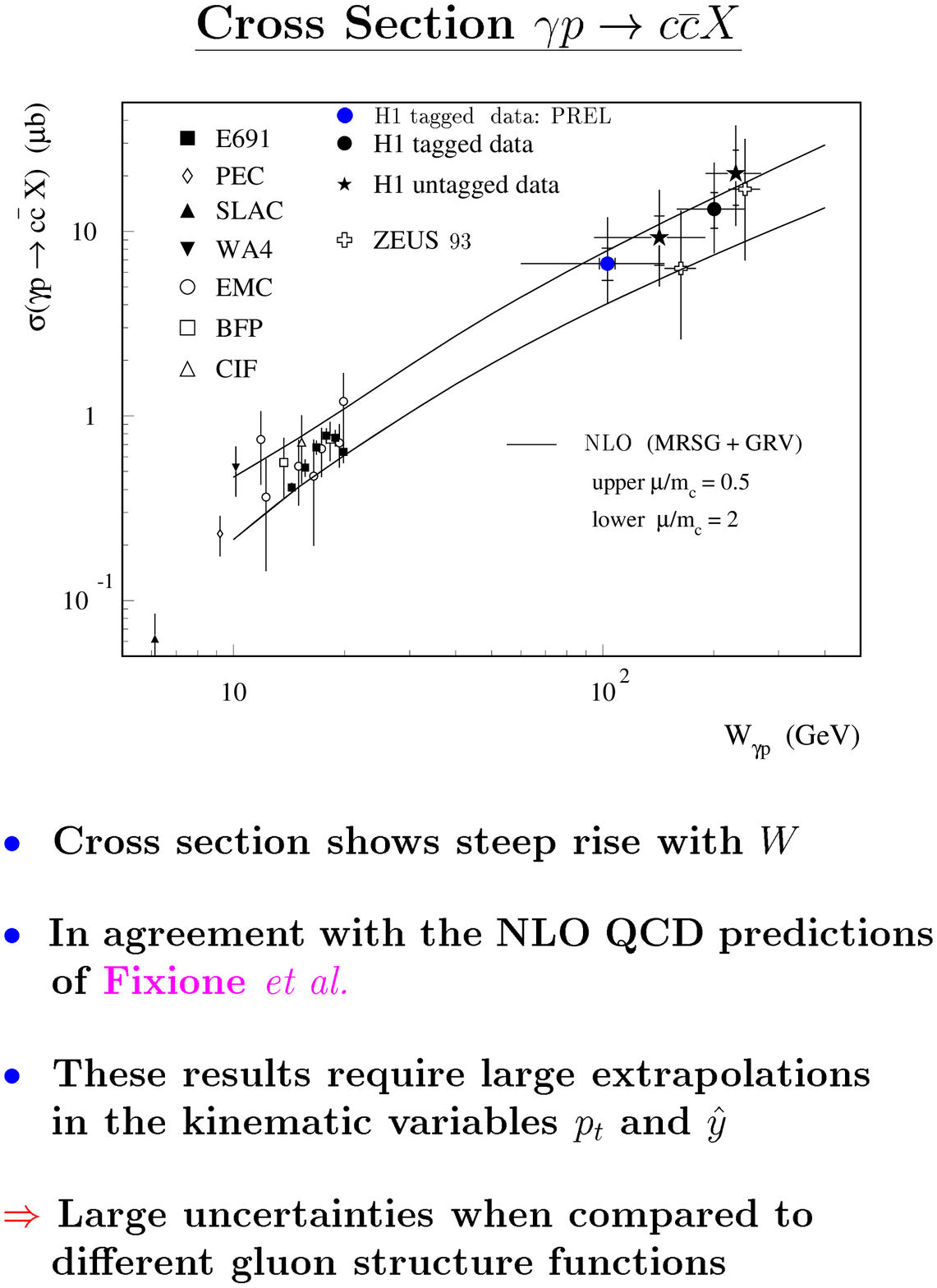,bbllx=60,bblly=380,
bburx=505,bbury=725,clip=true,scale=0.475}
\caption{\label{fig-sigmavsW}
The total cross-section for charm photoproduction, plotted as a function
of $W$. Error bars include uncertainties due to the model--dependent
extrapolation outside the experimental acceptance. Data points from H1
are plotted with inner error bars showing the contribution from
statistical and experimental systematic errors only.}
\end{figure}
From the number of reconstructed $\Dstar$s, the experiments measure the
visible cross-section for the reaction $e^+p \longrightarrow \Dstar X$
within the kinematic cuts applied, by correcting for acceptance and for
the combined branching ratio to the decay channels detected. In order to
extract the total charm production cross-section, the visible fraction
falling within the cuts must be estimated from Monte Carlo. A further
correction must be applied for the probability
$p_{c\longrightarrow{\Dstar}^+}$ for the charm quark to fragment to a
$\Dstar$ meson. The Monte Carlo extrapolation to the full $\Dstar$
kinematic region is subject to large uncertainties, arising mainly from
the choice of structure functions and of the value of $m_c$ used in the
calculations, and this uncertainty is reflected in the cross-section
error bars as plotted in Figure~\ref{fig-sigmavsW}. The data show an
order of magnitude rise in the integrated cross-section relative to
lower energy photoproduction measurements, in general agreeement with
the NLO QCD calculations of~\cite{bib-massivec}.

\begin{figure}[htb]\centering
\epsfig{file=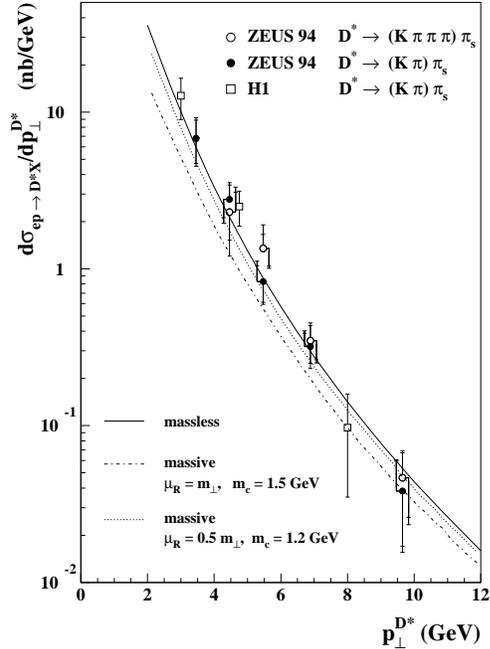,scale=0.4}
\caption{\label{fig-sigmadiff}
The $\Dstar$ $p_t$ spectrum in photoproduction compared with QCD
calculations.}
\end{figure}
The calculations of differential cross-sections, covering the
$(p_t,\eta,W)$ acceptance of the measurements, are much less sensitive
than the total cross-section to the uncertainties mentioned above. When
these distributions are plotted, the data are found to lie
systematically above the predictions of~\cite{bib-massivec}. The
predictions are based on an approach to performing NLO calculations
where only light quarks are assumed to be active flavours in the
structure functions of the proton and photon. Alternative calculations
have been performed\cite{bib-masslessc1,bib-masslessc2} where charm
flavour excitation is allowed to contribute in leading order, giving
generally higher predicted cross-sections. These are due principally to
the process $cg \longrightarrow cg$, where the hard scattering is
between a charm quark from the photon and a gluon from the proton. The
two types of calculation are described as massive and massless charm
approaches, respectively. Figure~\ref{fig-sigmadiff} shows the measured
$p_t$ spectrum from both experiments, together with the massive charm
predictions of \cite{bib-massivec} using two different values for the
charm quark mass $m_c$, and the massless charm predictions of
\cite{bib-masslessc1}. The massless charm approach is seen to give a
better description of the data. The calculations of
\cite{bib-masslessc2}, not plotted in Figure~\ref{fig-sigmadiff}, are
based on a different treatment of the heavy quark fragmentation and also
agree well.

\section{OPEN CHARM IN DEEP INELASTIC SCATTERING}
\label{sect-DIS}

Measurements of charm meson production in DIS events have been used by
H1\cite{bib-H1-DIS} and Zeus\cite{bib-ZEUS-DIS} to evaluate the charm
contribution to the proton structure function $F_2(x,Q^2)$. This
contribution $\Fcharm(x,Q^2)$ is defined by
\begin{equation}\frac{\partial^2 \sigma^{c\overline{c}}}{\partial x
\partial Q^2} = \frac{2\pi\alpha^2}{xQ^4} \left( 1 + \left( 1-y
\right)^2 \right) \Fcharm(x,Q^2) \end{equation}
where $\sigma^{c\overline{c}}$ is the total charm production
cross-section. Here the contribution due to longitudinally polarised
virtual photons, which is expected to be small, has been neglected. The
measurements of $\Fcharm$ are also discussed in the contribution on
structure functions to this conference\cite{bib-Zhokin}.

The experiments perform their analyses starting from samples selected by
the standard cuts for DIS event samples, principally by requiring the
detection of an energetic scattered positron. ZEUS use events in the
range $5<Q^2<100\,\mbox{GeV}^2$ and H1 require
$10<Q^2<100\,\mbox{GeV}^2$. The $\Dstar$ decay
channel~(\ref{eq-Dstdecay}) is again used as the signal for charm
production by both experiments, while H1 have additionally analysed
measurements of inclusive $D^0$ production.

\begin{figure}[htb]\centering
\epsfig{file=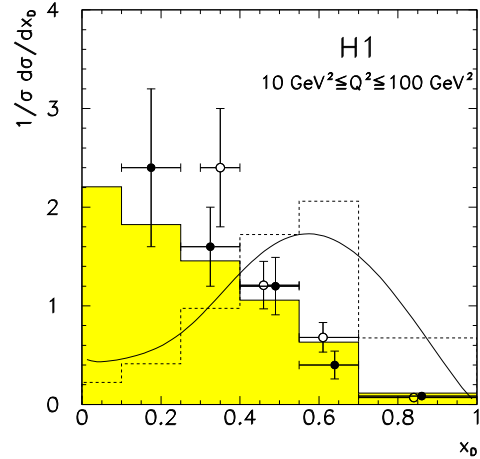,scale=0.375}
\caption{\label{fig-H1xddist}
Distribution of $D^0$ fractional momentum $x_D$ in DIS measured by H1
(points) compared with expectations from a PGF Monte Carlo
(AROMA{\protect\cite{bib-AROMA}}, shaded histogram) and from charm quark
sea contributions (dashed histogram and curve).}
\end{figure}
In order to throw light on the underlying production mechanism, the
fractional momentum $x_D$ of the charm mesons in the $\gamma^{\ast}p$
frame is studied. This quantity is defined by H1 as
\begin{equation} x_D = \frac{\left|p_D^{\ast}\right|}
{\left|p_p^{\ast}\right|} = 2\frac{\left|p_D^{\ast}\right|}{W}
\end{equation}
where $p_D^{\ast}$ and $p_p^{\ast}$ are the momenta in this frame of the
$D^0$ and the proton, respectively. The distribution is shown in
Figure~\ref{fig-H1xddist}. ZEUS present a similar distribution using the
momentum of the $\Dstar$ instead of the $D^0$. If the charm particles
are produced via photon-gluon fusion (PGF), a $c\overline{c}$ pair is
recoiling against the proton remnant in the $\gamma^{\ast}p$ frame. In
this case the $x_D$ distribution is expected to peak below $x_D=0.5$, as
observed. In the flavour excitation process where the virtual photon
interacts with a charm quark from the sea, the distribution is expected
to be shifted towards larger values of $x_D$. As shown in
Figure~\ref{fig-H1xddist}, the data are consistent with expectations
from a pure PGF Monte Carlo. The distributions of the observed events in
$p_t^D$, $\eta^D$, $W$ and  $Q^2$ are also found to be in good agreement
with predictions.

\begin{figure}[htb]\centering
\epsfig{file=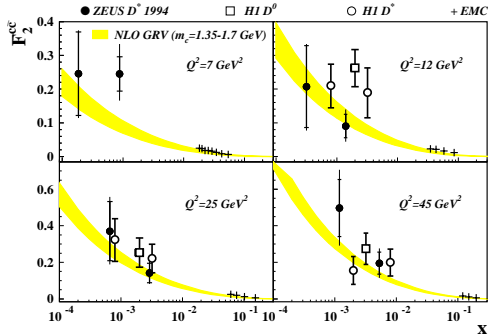,bbllx=0,bblly=200,
bburx=600,bbury=600,clip=true,scale=0.35}
\caption{\label{fig-F2cmeas}
Measurements of $\Fcharm$ from 1994 HERA data, plotted as a function of
$x$ in 4 bins of $Q^2$. Also shown are fixed target measurements from
EMC. The shaded bands are NLO QCD predictions for a range of charm quark
masses.}
\end{figure}
The measurements of $\Fcharm$ are shown in Figure~\ref{fig-F2cmeas},
together with results from the EMC\cite{bib-EMCF2c} fixed target
experiment at higher $x$. For the low values of $x$ accessible at HERA,
$x \sim 10^{-3}$, it is found that $\Fcharm$ is around 25\% of the total
$F_2$. Also shown in Figure~\ref{fig-F2cmeas} is the range of
predictions from an NLO calculation\cite{bib-Harris} using the
GRV\cite{bib-GRV} gluon distribution, for charm quark masses between
$1.35 < m_c < 1.7\,\mbox{GeV}$. The lowest value of the mass corresponds
to the upper boundary of the shaded region in the figure.

The NLO QCD calculation of~\cite{bib-Harris} includes only the PGF
diagram as the source of charm production in DIS. This approach is
therefore similar to the massive charm calculations for photoproduction
discussed in section~\ref{sect-gprod}. There it was seen that such an
approach gives cross-section predictions which tend to be lower than the
photoproduction data, while for DIS the agreement is good within the
limited statistical precision of the 1994 data samples.

Again for DIS, there exist alternative treatments within the framework
of perturbative QCD, which treat the charm quark as an active parton
with its own distribution function $c(x,Q^2)$. The charm distribution is
driven by evolution from the gluon density, being zero below some
threshold $Q^2$ value of order $m_c^2$. These treatments are expected to
give improved results for large $Q^2$ phenomena. A number of
authors\cite{bib-Collins,bib-CTEQ,bib-MRRS,bib-96-258} have recently
discussed ways to implement such schemes so that the behaviour at $Q^2
\sim m_c^2$ is essentially given by PGF, to give a description of charm
production valid for all scales. Such analyses predict larger values of
$\Fcharm$ at high $Q^2$ than expected for pure PGF, a prediction which
should be testable with the higher statistics 1995-6 HERA data now being
analysed.

\section{SUMMARY}

Charm production at HERA in real and virtual photon--proton collisions
provides an excellent laboratory for detailed tests of perturbative QCD
models. This is an area of considerable experimental and theoretical
activity. Models of photoproduction are in reasonable agreement with
data at current levels of precision, with some suggestion of the need
for a contribution due to ``active charm'' in the photon. Charm
production is found to contribute a substantial fraction of the proton
structure function at small $x$, as expected given the observed steep
rise of the gluon distribution. Precise measurements of $\Fcharm(x,Q^2)$
will provide a useful independent constraint on fits to parton
distributions, and a check of different approaches to calculations of
heavy quark prodution in perturbative QCD.


\begin{thebibliography}{99}
\bibitem{bib-Meyer} A.\ Meyer, these proceedings.
\bibitem{bib-H1-gprod} H1 collaboration, S.\ Aid {\em et al.}, Nucl.\
Phys.\ {\bfseries B472} (1996), 32.
\bibitem{bib-ZEUS-gprod} ZEUS collaboration, J.\ Breitweg {\em et al.},
preprint hep-ex/9704011, accepted by Phys.\ Lett.\ {\bfseries B}.
\bibitem{bib-massivec} S.\ Frixione {\em et al.}, Nucl.\ Phys.\
{\bfseries B454} (1995), 3, and further references therein.
\bibitem{bib-masslessc1} B.A.\ Kniehl {\em et al.}, Phys.\ Lett.\
{\bfseries B356} (1995), 539.
\bibitem{bib-masslessc2} M.\ Cacciari, M.\ Greco, Z.\ Phys.\ {\bfseries
C69} (1996), 459.
\bibitem{bib-H1-DIS} H1 collaboration, C.\ Adloff {\em et al.},
Z.\ Phys.\ {\bfseries C72} (1996), 593.
\bibitem{bib-ZEUS-DIS} ZEUS collaboration, J.\ Breitweg {\em et al.},
preprint hep-ex/9706009, submitted to Z.\ Phys.\ {\bfseries C}.
\bibitem{bib-Zhokin} A.\ Zhokin, these proceedings.
\bibitem{bib-AROMA} G.\ Ingelman, J.\ Rathsman, G.A.\ Schuler, preprint
hep-ph/9605285.
\bibitem{bib-EMCF2c} EMC collaboration, J.J.\ Aubert {\em et al.},
Nucl.\ Phys.\ {\bfseries B213} (1983), 31.
\bibitem{bib-Harris} B.W.\ Harris, J.\ Smith, preprint hep-ph/9706334.
\bibitem{bib-GRV} M.\ Gl\"{u}ck, E.\ Reya, A.\ Vogt, Z.\ Phys.\
{\bfseries C67} (1995), 27. 
\bibitem{bib-Collins} M.A.G.\ Aivazis {\em et al.}, Phys.\ Rev.\
{\bfseries D50} (1994), 3102
\bibitem{bib-CTEQ} H.L.\ Lai, W.K.\ Tung, Z.\ Phys.\ {\bfseries C74}
(1997), 463
\bibitem{bib-MRRS} A.D.\ Martin {\em et al.}, preprint hep-ph/9612449.
\bibitem{bib-96-258} M.\ Buza {\em et al.}, preprint hep-ph/9612398.
\end{thebibliography}
\end{document}